\documentclass{article}
\usepackage{amssymb}
\usepackage{amsfonts}
\usepackage{amsmath}

\newtheorem{theorem}{Theorem}
\newtheorem{acknowledgement}[theorem]{Acknowledgement}

\input{tcilatex}

\begin{document}

\title{Non-Hermitian Hamiltonian versus $E=0$ localized states}
\author{S. Habib Mazharimousavi \\
Department of Physics, Eastern Mediterranean University, \\
G Magusa, north Cyprus, Mersin 10,Turkey\\
E-mail: habib.mazhari@emu.edu.tr}
\maketitle

\begin{abstract}
We analyze the zero energy solutions,\ of a two dimensional system which
undergoes a non-radial symmetric, complex potential $V\left( r,\phi \right) $%
. By virtue of the coherent states concept, the localized states are
constructed, and the consequences of the imaginary part of the potential are
found both analytically and schematically.
\end{abstract}

\section{Introduction}

Since the early years of quantum mechanics the exact solvability of quantum
mechanical models have attracted much attention. Some exactly solvable
models have already become typical standard examples in quantum mechanical
textbooks. However, it was believed that the reality of the spectra of the
Hamiltonians, describing quantum mechanical models, is necessarily
attributed to their Hermiticity. It was the non-Hermitian $PT$-symmetric
Hamiltonians proposed by Bender and Boettcher [1] that relaxed the
Hermiticity condition as a necessity for the reality of the spectrum [1--7].
Herein, $P$ denotes the parity ($PxP=-x$) and the anti-linear operator $T$
mimics the time reflection ($TiT=-i$). Recently, Mostafazadeh [8] has
introduced a broader class of non-Hermitian pseudo-Hermitian Hamiltonians (a
generalization of $PT$-symmetric, therefore). In these settings [8--19], a
Hamiltonian $H$ is pseudo-Hermitian if it obeys the similarity
transformation: $\eta H\eta ^{-1}=H^{\dag }$ where $\eta $ is a Hermitian
invertible linear operator. On the other hand, the study of the $E=0$ bound
states have found many applications in various fields [20-24]. Long ago,
Daboul and Nieto [25,26] have discovered that, an attractive radial power
low potential, $V(r)\sim r^{-\nu }$ for $\nu <-2$ and $\nu >2,$ passes
through the $E=0$ normalizable solutions. More recently Makowski and G\'{o}%
rska, established the classical correspondence localized states of a system
with zero energy and a general form of power low potentials [27]. In their
work, it was shown that the classical trajectories of the particle precisely
matched with the localized quantum states.

In this work we attempt to understand the consequences of adding an
imaginary term to a potential whose zero energy level passes through the $%
E=0 $ normalizable bound state solution (namely $V(r)=-\Gamma r^{-4})$. We
believe that this kind of study is necessary, to find relations, if any
between the non-Hermitian quantum mechanics and the classical mechanics.

The organization of our paper is as follows: In Sec. (2) we give an
analytical solution to the Schr\"{o}dinger equation of a zero energy
particle under our chosen complex potential. We continue in Sec. (3) by
adapting a closed form of the localized states from literature and then we
set up and plot the classically equivalent coherent states of the system. We
conclude our paper with Sec. (4).

\section{Analytic solution of the Schr\"{o}dinger equation}

Two dimensional Schr\"{o}dinger equation for a zero energy particle under a
complex, non-radial symmetric potential is given by%
\begin{equation}
\left[ -\frac{\hslash ^{2}}{2m}\left( \frac{1}{r}\frac{\partial }{\partial r}%
\left( r\frac{\partial }{\partial r}\right) +\frac{1}{r^{2}}\frac{\partial
^{2}}{\partial \phi ^{2}}\right) +V\left( r,\phi \right) \right] \psi \left(
r,\phi \right) =0
\end{equation}%
where 
\begin{equation}
V\left( r,\phi \right) =-\frac{\Gamma }{r^{4}}-\frac{\Lambda }{r^{2}}%
e^{i\phi }
\end{equation}%
and for our purpose, $\Gamma $ and $\Lambda $ are some non-negative real
constants. Before we go further, finding out the symmetric properties of $%
V\left( r,\phi \right) $ and consequently the Hamiltonian of the system may
give some connections between this potential and the well known $\mathcal{PT}
$-symmetric type potentials which are studied in the literature.

Let us introduce an operator $\Theta $ which is defined as 
\begin{equation}
\Theta :i\rightarrow -i,\phi \rightarrow 2\pi -\phi
\end{equation}%
in which $i=\sqrt{-1}$ and $\phi $ is the usual azimuthal angle. One can
easily show that $\Theta $ is a non-Hermitian, invertible operator whose
inverse is given by%
\begin{equation}
\Theta ^{-1}=\Theta
\end{equation}%
and it can be decomposed into the two other operators $\mathcal{T}$ and $\Pi 
$ such that%
\begin{equation}
\Theta =\Pi \mathcal{T}
\end{equation}%
in which the definition of these operators are given by%
\begin{eqnarray}
\Pi &:&\phi \rightarrow 2\pi -\phi \\
\mathcal{T} &:&i\rightarrow -i.  \notag
\end{eqnarray}

$\Pi $ is Hermitian and invertible such that $\Pi ^{-1}=\Pi =\Pi ^{\dag },$
and $\mathcal{T}$ \ is the usual time reversal operator. It is remarkable to
observe that the Hamiltonian of the particle under the potential (2) is $\Pi 
\mathcal{T}$-symmetric which means%
\begin{equation}
H=H^{\Pi \mathcal{T}}.
\end{equation}

Also it is said that the $\Pi \mathcal{T}$-symmetry of a Hamiltonian $H$ is
unbroken if all of the eigenfunctions of $H$ are simultaneously
eigenfunctions of $\Pi \mathcal{T}$. It is easy to show that if the $\Pi 
\mathcal{T}$-symmetry of a Hamiltonian $H$ is unbroken, then the spectrum of 
$H$ is real (to see the proof, one may see [30])$.$

We come back to the equation (1) and as usual, we take%
\begin{equation}
\psi \left( r,\phi \right) =R\left( r\right) \Phi \left( \phi \right) 
\end{equation}%
and after making substitution, in Eq. (1) we get a set of two equations as 
\begin{gather}
\left[ \frac{1}{r}\frac{\partial }{\partial r}\left( r\frac{\partial }{%
\partial r}\right) -\frac{l^{2}}{r^{2}}+\frac{2m\Gamma }{\hslash ^{2}r^{4}}%
\right] R\left( r\right) =0 \\
and  \notag \\
\frac{\partial ^{2}\Phi \left( \phi \right) }{\partial \phi ^{2}}+\frac{%
2m\Lambda }{\hslash ^{2}}e^{i\phi }\Phi \left( \phi \right) =-l^{2}\Phi
\left( \phi \right) 
\end{gather}%
where $l$ is a constant to be identified.

By introducing a new dimensionless variable $\rho $ as 
\begin{equation}
\rho =\frac{r}{a_{\circ }}
\end{equation}%
where $a_{\circ }$ is a positive constant, the Eqs. (9) and (10) become%
\begin{gather}
\left[ \frac{1}{\rho }\frac{d}{d\rho }\left( \rho \frac{d}{d\rho }\right) -%
\frac{l^{2}}{\rho ^{2}}+\frac{\gamma ^{2}}{\rho ^{4}}\right] R\left(
r\right) =0 \\
and  \notag \\
\frac{d^{2}\Phi \left( \phi \right) }{d\phi ^{2}}+\lambda ^{2}e^{i\phi }\Phi
\left( \phi \right) =-l^{2}\Phi \left( \phi \right) .
\end{gather}%
in which%
\begin{eqnarray}
\gamma ^{2} &=&\frac{2m\Gamma }{\hslash ^{2}a_{\circ }^{2}} \\
&&and  \notag \\
\lambda ^{2} &=&\frac{2m\Lambda }{\hslash ^{2}}.
\end{eqnarray}

In the angular part of the Schr\"{o}dinger equation we change the variable,
and introduce 
\begin{equation}
\chi =e^{i\phi }
\end{equation}%
hence the Eq. (13) reads%
\begin{equation}
\chi ^{2}\frac{d^{2}\Phi \left( \chi \right) }{d\chi ^{2}}+\chi \frac{d\Phi
\left( \chi \right) }{d\chi }-\left( l^{2}+\lambda ^{2}\chi \right) \Phi
\left( \chi \right) =0.
\end{equation}%
This is the modified Bessel ODE, such that its complete solution is well
known as%
\begin{eqnarray}
\Phi \left( \chi \right) &=&C_{1}I_{2l}\left( 2\lambda \sqrt{\chi }\right)
+C_{2}K_{2l}\left( 2\lambda \sqrt{\chi }\right)  \notag \\
&&or \\
\Phi \left( \phi \right) &=&C_{1}I_{2l}\left( 2\lambda e^{i\phi /2}\right)
+C_{2}K_{2l}\left( 2\lambda e^{i\phi /2}\right)  \notag
\end{eqnarray}%
in which $I_{\nu }\left( z\right) $ and $K_{\nu }\left( z\right) $ are the
modified Bessel functions.

This solution should satisfy the following boundary condition

\begin{equation}
\Phi \left( \phi \right) =\Phi \left( \phi +2\pi \right)
\end{equation}%
or equivalently 
\begin{gather}
C_{1}I_{2l}\left( 2\lambda e^{i\phi /2}\right) +C_{2}K_{2l}\left( 2\lambda
e^{i\phi /2}\right) = \\
C_{1}I_{2l}\left( 2\lambda e^{i\left( \phi +2\pi \right) /2}\right)
+C_{2}K_{2l}\left( 2\lambda e^{i\left( \phi +2\pi \right) /2}\right) . 
\notag
\end{gather}

Since $I_{\nu }\left( z\right) $ and $K_{\nu }\left( z\right) $ are two
independent solutions of the modified Bessel ODE , a class of solution is
possible when we put $C_{2}=0.$ Therefore we get 
\begin{equation}
I_{2l}\left( 2\lambda e^{i\phi /2}\right) =I_{2l}\left( 2\lambda e^{i\left(
\phi +2\pi \right) /2}\right) .
\end{equation}

In accordance with the following property of the modified Bessel functions
[28] 
\begin{equation}
I_{\nu }\left( ze^{m\pi i}\right) =e^{m\pi \nu i}I_{\nu }\left( z\right)
\end{equation}%
where $z$ is a complex variable, $m$ is an integer and $\nu $ is a real
number, one can choose 
\begin{equation}
z=2\lambda e^{i\phi /2}
\end{equation}%
and $m=1$ in Eq. (22) to get 
\begin{equation}
I_{2l}\left( ze^{i\pi }\right) =e^{i2l\pi }I_{2l}\left( z\right) .
\end{equation}%
This equality is valid for all $z$ and $l$ in their domains, but if one
considers $l$ to be an integer, we will get%
\begin{equation}
I_{2l}\left( ze^{i\pi }\right) =I_{2l}\left( z\right)
\end{equation}%
which is equivalent to Eq. (22). Therefore $l$ is found to be an integer i.e.%
\begin{equation}
l=0,\pm 1,\pm 2,...
\end{equation}

Of course as a different possibility, one can choose $C_{1}=0$ to find a
different class of solution but the following property [28] 
\begin{equation}
K_{\nu }\left( ze^{im\pi }\right) =e^{-m\pi \nu i}K_{\nu }\left( z\right)
-\pi i\sin m\nu \pi \csc \nu \pi I_{\nu }\left( z\right)
\end{equation}%
it comes to our setting as%
\begin{equation}
K_{2l}\left( ze^{i\pi }\right) =e^{-\pi 2li}K_{2l}\left( z\right) -\pi i\sin
2l\pi \csc \nu \pi I_{2l}\left( z\right)
\end{equation}%
which obviously does not admit any solution.

As a result, the solutions of the angular part of the Schr\"{o}dinger
equation can be written explicitly as%
\begin{eqnarray}
\Phi _{l}\left( \phi \right) &=&C_{l\lambda }I_{2l}\left( 2\lambda e^{i\phi
/2}\right) \\
l &=&0,1,2,...  \notag
\end{eqnarray}%
where, since $I_{-n}\left( z\right) =I_{n}\left( z\right) ,$ we just
consider $l$ to be non negative, and $C_{l\lambda }$ are the normalization
constants given by 
\begin{equation}
C_{l\lambda }=\sqrt{\frac{1}{\int_{0}^{2\pi }\left\vert \Phi _{l}\left( \phi
\right) \right\vert ^{2}d\phi }}=\sqrt{\frac{1}{\int_{0}^{2\pi }\left\vert
I_{2l}\left( 2\lambda e^{i\phi /2}\right) \right\vert ^{2}d\phi }}.
\end{equation}

\bigskip One may note that $l$ still can be interpreted as the angular
quantum number, since 
\begin{eqnarray}
&<&\hat{L}\underset{\Phi _{l}}{>}=<\Phi _{l}|-i\hslash \frac{\partial }{%
\partial \phi }|\Phi _{l}>= \\
&<&C_{l\lambda }I_{2l}\left( 2\lambda e^{i\phi /2}\right) |-i\hslash \frac{%
\partial }{\partial \phi }|C_{l\lambda }I_{2l}\left( 2\lambda e^{i\phi
/2}\right) >=  \notag \\
-i\hslash \left\vert C_{l\lambda }\right\vert ^{2} &<&I_{2l}\left( 2\lambda
e^{i\phi /2}\right) |i\left[ \lambda e^{i\phi /2}I_{2l+1}\left( 2\lambda
e^{i\phi /2}\right) +lI_{2l}\left( 2\lambda e^{i\phi /2}\right) \right]
>=l\hslash .  \notag
\end{eqnarray}

The radial part of the Schr\"{o}dinger equation, can be considered as the
Bessel ODE if one defines 
\begin{equation}
\xi =\frac{1}{\rho }
\end{equation}%
and therefore the Eq. (12) becomes%
\begin{equation}
\xi ^{2}\frac{d^{2}}{d\xi ^{2}}R\left( \xi \right) +\xi \frac{d}{d\xi }%
R\left( \xi \right) +\left( \gamma ^{2}\xi ^{2}-l^{2}\right) R\left( \xi
\right) =0
\end{equation}%
which admits a complete solution 
\begin{equation}
R_{l\gamma }\left( \xi \right) =C_{1}J_{l}\left( \gamma \xi \right)
+C_{2}Y_{l}\left( \gamma \xi \right) .
\end{equation}

A physical,\ normalizable solution which for $l>1$ corresponds to the bound
state is given by [25,26]%
\begin{eqnarray}
R_{l\gamma }\left( \xi \right) &=&N_{l\gamma }J_{l}\left( \gamma \xi \right)
\\
l &=&2,3,...  \notag
\end{eqnarray}%
in which $N_{l\gamma }$ are normalization constants given by%
\begin{equation}
N_{l\gamma }=\frac{2}{a_{\circ }\gamma }\sqrt{\frac{\left( l+1\right) !}{%
\left( l-2\right) !}}.
\end{equation}

Now we are ready to write the complete solution of the Schr\"{o}dinger
equation (i.e. wave function), by using the Eqs. (29) and (35) as%
\begin{equation}
\psi _{l,\gamma ,\lambda }\left( r,\phi \right) =C_{l\lambda }\frac{2}{%
a_{\circ }\gamma }\sqrt{\frac{\left( l+1\right) !}{\left( l-2\right) !}}%
I_{2l}\left( 2\lambda e^{i\phi /2}\right) J_{l}\left( \frac{\gamma a_{\circ }%
}{r}\right) .
\end{equation}

We notice that, with $l>1$, the only complex part of $\psi _{l,\gamma
,\lambda }\left( r,\phi \right) $ is the modified Bessel function. As a
matter of fact, the effect of $\Theta $ (introduced in Eq. (3)) on $\psi
_{l,\gamma ,\lambda }\left( r,\phi \right) $ is equivalent to the effect of $%
\Theta $ on $I_{2l}\left( 2\lambda e^{i\phi /2}\right) .$ Therefore, by
using the expansion form of the modified Bessel function, this equation can
be written as%
\begin{equation}
\psi _{l,\gamma ,\lambda }\left( r,\phi \right) =C_{l\lambda }\frac{2}{%
a_{\circ }\gamma }\sqrt{\frac{\left( l+1\right) !}{\left( l-2\right) !}}%
J_{l}\left( \frac{\gamma a_{\circ }}{r}\right) \underset{s=0}{\overset{%
\infty }{\sum }}\frac{\lambda ^{2\left( s+l\right) }}{\Gamma \left(
s+2l+1\right) s!}e^{i\phi \left( s+l\right) }
\end{equation}%
in which,clearly this is invariant under $\Theta $ (i.e., $\Theta $ $\psi
_{l,\gamma ,\lambda }\left( r,\phi \right) =\psi _{l,\gamma ,\lambda }\left(
r,\phi \right) $).

\subsection{A realized approach to the problem}

In this section we will consider the following radial symmetric real
potentials%
\begin{equation}
V_{\pm }\left( r\right) =-\frac{\Gamma }{r^{4}}\pm \frac{\Lambda }{r^{2}}
\end{equation}%
where $\Gamma $ and $\Lambda $ are some positive constants as before. One
should notice that the negative branch of the above potential is same as the
potential in Eq. (2) in an attractive form, and the positive branch of it is
same but in a repulsive form. The Schr\"{o}dinger equation (1), after the
usual separation method and change of variable, with \ the potential (38)
comes to a set of two separated equations as%
\begin{gather}
\left[ \frac{1}{\rho }\frac{d}{d\rho }\left( \rho \frac{d}{d\rho }\right) -%
\frac{\tilde{l}^{2}}{\rho ^{2}}+\frac{\gamma ^{2}}{\rho ^{4}}\right] R\left(
\rho \right) =0 \\
and  \notag \\
\frac{d^{2}\Phi \left( \phi \right) }{d\phi ^{2}}=-l^{2}\Phi \left( \phi
\right)
\end{gather}%
where 
\begin{equation*}
\tilde{l}^{2}=l^{2}\pm \lambda ^{2}
\end{equation*}%
in which the positive (negative) sign is related to the +$\frac{\Lambda }{%
r^{2}}$ ($-\frac{\Lambda }{r^{2}}$)$,$and the other factors are defined as
before. One can easily show that, the final solution of the Schr\"{o}dinger
equation with the potentials (38) can be written as 
\begin{equation}
\psi _{l,\gamma ,\lambda }\left( r,\phi \right) =\frac{1}{\sqrt{2\pi }}\frac{%
2}{a_{\circ }\gamma }\sqrt{\frac{\left( \tilde{l}+1\right) !}{\left( \tilde{l%
}-2\right) !}}e^{il\phi }J_{\tilde{l}}\left( \frac{1}{r}\right)
\end{equation}%
in which $\tilde{l}$ must be greater then $1$ to have bound states. In the
sequel we will use the closed forms of the infinite number of degenerate
wave functions (i.e., these states have same energy equal to zero),
presented in the Eq.s (37) and (41) to construct the classical
correspondence localized states.

\section{Zero energy localized states}

In references [25-27] it was shown that the trajectory of a classical
particle which experiences a real potential in the form of the $\Gamma -$%
part of the potential considered in this work (2) (i.e. $-\frac{\Gamma }{%
r^{4}}$) is given by 
\begin{equation}
\left\{ 
\begin{array}{l}
x=\frac{a}{2}(1+\cos \left( \phi -\phi _{\circ }\right) ) \\ 
y=\frac{a}{2}\sin \left( \phi -\phi _{\circ }\right)%
\end{array}%
\right.
\end{equation}%
which represents a circle with radius $a/2$ (i.e., if one set $\phi _{\circ
}=0,$ this becomes $\left( x-a/2\right) ^{2}+y^{2}=\left( a/2\right) ^{2}$)
such that 
\begin{equation}
a=\sqrt{2m\Gamma /L^{2}}
\end{equation}%
and $L$ is the conserved angular momentum of the particle. We notice that,
the classical correspondence localized state (see [27] and the references
therein) of the potential (2) while $\Lambda \rightarrow 0$ must have
probability peak in accordance with the classical trajectory (i.e., a circle
with radius a/2 as implied by Eq. (42) [27] ). Our aim in the following is
to see the effect of the $\Lambda -$part of the potential (2) on the shape
of the classical correspondence localized states. To this end, first we find
the localized states of the original potential (2) in a closed analytical
form, then we follow similarly but for the case when the potential is in the
forms of Eq. (38).\ 

An available method to derive the corresponding localized states by using
the solutions of the Schr\"{o}dinger equation given in the previous
sections, is based on the concept of deformed oscillator algebras [27,29].
Therefore a suitable explicit form of the localized state over the infinite
number of degenerate states ( with $E=0$) $\psi _{l\gamma \lambda }\left(
r,\phi \right) $ reads (see [27,29] and the references therein)%
\begin{equation}
\Psi _{N}=\frac{1}{\sqrt{2\pi }\left( 1+\left\vert \tau \right\vert
^{2}\right) ^{N/2}}\overset{N}{\underset{k=0}{\sum }}\binom{N}{k}^{1/2}\tau
^{k}\psi _{k,\gamma ,\lambda }\left( r,\phi \right)
\end{equation}%
in which $k=l-2$ and $\tau =Ae^{i\theta _{\circ }},$ where $A$ and $\theta
_{\circ }$ are some real constants$.$

\subsection{Results}

The classical correspondence localized states of a particle undergoes the
potentials (2) and (38), by choosing $N=7$ may be written as%
\begin{equation}
\Psi _{7}=\left\{ 
\begin{array}{ll}
\frac{1}{8\sqrt{\pi }}\overset{7}{\underset{k=0}{\sum }}C_{\left( 2+k\right)
\lambda }\sqrt{\frac{\binom{7}{k}\left( k+3\right) !}{k!}}I_{\left(
4+2k\right) }\left( 2\lambda e^{i\frac{\phi }{2}}\right) J_{\left(
2+k\right) }\left( \frac{1}{r}\right) & V=V_{c} \\ 
\frac{1}{16\pi }\overset{7}{\underset{k=0}{\sum }}\sqrt{\frac{\binom{7}{k}%
\left( \tilde{l}+1\right) !}{\left( \tilde{l}-2\right) !}}e^{il\phi }J_{%
\tilde{l}}\left( \frac{1}{r}\right) & V=V_{\pm }%
\end{array}%
\right. .
\end{equation}%
in which $\gamma a_{\circ }$ and $A$ are set to be one and for convenience $%
V_{c}$ and $V_{\pm }$ refer to the potentials (2) and (38) respectively.
Within the Fig.s (1) to (7) some density plot of $\left\vert \Psi
_{7}\right\vert ^{2}$ are given in terms of different values of $\lambda .$%
In the Fig. (1) we plot $\left\vert \Psi _{7}\right\vert ^{2}$ with $\lambda
=0$ (i.e., $V_{c}$ = $V_{\pm }=-\frac{\Gamma }{r^{4}})$ and the classical
trajectory of the particle ( this figure was reported in the Ref.[27]). In
the Fig.s (2) to (4), (a), (b) and (c) refer to the potentials $V_{-},$ $%
V_{+}$ and $V_{c},$ respectively. In the Fig.s (5) and (6), (a) refers to $%
V_{+}$ and (b) refers to $V_{c}.$ Finally Fig. (7) refers to $V_{c}.$

\subsection{Behavior of the complexified modified Bessel functions}

In this section we give a short description of the behavior of the
complexified modified Bessel functions $I_{2l}(z)$ to explain why with $%
V=V_{c}$ the plot of the probability density with large values of $\lambda $
are very localized around $\phi =0.$ It is not difficult to see that, in the
case of $V=V_{c}$ the $\Lambda -$part of the original potential (2) goes
into the $\phi $-part of the Schr\"{o}dinger equation and therefore the
entire effect of this term appears in the $\phi $-part of the wave function.
Therefore the contribution of $C_{l\lambda }I_{2l}\left( 2\lambda e^{i\phi
/2}\right) $ in the final form of the wave function, instead of the usual $%
\phi -$part in the wave functions (i.e.,$\frac{1}{\sqrt{2\pi }}e^{il\phi }$)
of the case of $V=V_{\pm }$ is the reason of the great localization about
small $\phi $. Let us write 
\begin{equation}
I_{2l}\left( 2\lambda e^{i\phi /2}\right) =I_{2l}\left( 2\lambda \cos \left( 
\frac{\phi }{2}\right) +i2\lambda \sin \left( \frac{\phi }{2}\right) \right)
\end{equation}%
which shows that the square root of the real (imaginary) part of the
potential directly goes through the real (imaginary) part of the argument of 
$I_{2l}\left( 2\lambda e^{i\phi /2}\right) .$ To see the behavior of $%
I_{2l}\left( 2\lambda e^{i\phi /2}\right) $ in terms of the real (imaginary)
part of its argument we rewrite the last equation as%
\begin{equation}
I_{2l}\rightarrow I_{2l}\left( 2\lambda _{1}\cos \left( \frac{\phi }{2}%
\right) +i2\lambda _{2}\sin \left( \frac{\phi }{2}\right) \right)
\end{equation}%
which in the limit of $\lambda _{1}=\lambda _{2}=\lambda $ turns out to be $%
I_{2l}\left( 2\lambda e^{i\phi /2}\right) .$ Figures (8),(9) and (10) show
that once $\lambda _{2}$ vanishes $\left\vert I_{2l}\right\vert ^{2}$ does
not change much, but once $\lambda _{1}$ becomes zero, $\left\vert
I_{2l}\right\vert ^{2}$ decreases strongly. Also once $\lambda _{1}$ takes a
larger value (it does not matter what is the value of $\lambda _{2}$), $%
\left\vert I_{2l}\right\vert ^{2}$ takes much higher value close to $\phi =0$
or $2\pi .$ We conclude therefore that, the imaginary part of the argument
of $I_{2l}$ (and imaginary part of the potential therefore) does not
contribute much in the localization of the $\left\vert \Psi _{7}\right\vert
^{2}$ around small $\phi .$ In contrast, the real part, and its $\phi -$%
dependent part (i.e., $\cos \phi $) causes such a great localization.

\section{Conclusion}

In conclusion, we concentrate ourselves and wish to comment on the figures.
Fig. (1) is our reference figure, i.e., the classical correspondence
localized state when the $\Lambda -$part of the potential vanishes. Fig.s
(2-4) clearly show that the effects of the $\Lambda -$part in the $V_{c}$
and $V_{+}$ cause a higher localization while in the $V_{-}$ we see a lower
localization. It is remarkable to observe that, the magnitude of the effects
(within these three cases) $V_{c}$ is greater than others. Fig.s (5,6) show
that as $\lambda $ takes large value the radius of the localized state
corresponding to the $V_{+}$ decreases while the $\phi $ distribution of the 
$\left\vert \Psi _{7}\right\vert ^{2}$ does not change. For the case of
potential $V_{c}$ the radii of the localized states are fixed while the $%
\phi $ distribution of the probability density $\left\vert \Psi
_{7}\right\vert ^{2}$ is changed so that the particle seems to be localized
more around $\phi =0.$ In the Fig. (7) we see the effect of the $\Lambda -$%
part in the $V_{c}$ as a great localization about $\phi =0.$ Obviously the
figures imply that in the case of $V=V_{c},$ the particle is localized
around $\phi =0,$ which is a direct consequence of the $\Lambda -$part of
the potential. We notice that, our approach to the problem is in a closed
analytical form, where all numerical results are based on the analytical
solutions. There is no need to comment that any other approach will
definitely lack the advantages of an exact analytical solution.

\begin{acknowledgement}
The author would like to thank Professor Mustafa Halilsoy and Professor Omar
Mustafa for fruitful discussions and useful comments.
\end{acknowledgement}

\begin{acknowledgement}
Also I would like to thank the anonymous referee for the valuable comments
and suggestions.
\end{acknowledgement}

\emph{Figure caption}

Figure (1): A plot of probability density $\left\vert \Psi _{N}\left( r,\phi
\right) \right\vert ^{2}$ for $N=7$ (i.e. $2\leq l\leq 9)$, $\theta _{\circ
}=0,$ $A=1$, $\epsilon =0$ (i.e. the $V(r,\phi )$ is real$)$ and $\lambda
=0.0$. This is the localized state corresponding to the classical trajectory
of a particle which experiences just the first term of the potential, i.e. $%
V(r,\phi )=-\Gamma /r^{4}$ and therefore it is the reference plot.

Figure (2): A plot of probability density $\left\vert \Psi _{N}\left( r,\phi
\right) \right\vert ^{2}$ for $N=7$ (i.e. $2\leq l\leq 9)$, $\theta _{\circ
}=0,$ $A=1$, $\lambda =0.1.$ Also (a), (b) and (c) are correspondence with $%
V(r,\phi )=V_{-}$,$V_{+}$ and $V_{c}$ respectively.

Figure (3): A plot of probability density $\left\vert \Psi _{N}\left( r,\phi
\right) \right\vert ^{2}$ for $N=7$ (i.e. $2\leq l\leq 9)$, $\theta _{\circ
}=0,$ $A=1$, $\lambda =0.5.$ Also (a), (b) and (c) are correspondence with $%
V(r,\phi )=V_{-}$,$V_{+}$ and $V_{c}$ respectively.

Figure (4): A plot of probability density $\left\vert \Psi _{N}\left( r,\phi
\right) \right\vert ^{2}$ for $N=7$ (i.e. $2\leq l\leq 9)$, $\theta _{\circ
}=0,$ $A=1$, $\lambda =1.$ Also (a), (b) and (c) are correspondence with $%
V(r,\phi )=V_{-}$,$V_{+}$ and $V_{c}$ respectively.

Figure (5): A plot of probability density $\left\vert \Psi _{N}\left( r,\phi
\right) \right\vert ^{2}$ for $N=7$ (i.e. $2\leq l\leq 9)$, $\theta _{\circ
}=0,$ $A=1$, $\lambda =5.$ Also (a) and (b) are correspondence with $%
V(r,\phi )=V_{+}$ and $V_{c}$ respectively.

Figure (6): A plot of probability density $\left\vert \Psi _{N}\left( r,\phi
\right) \right\vert ^{2}$ for $N=7$ (i.e. $2\leq l\leq 9)$, $\theta _{\circ
}=0,$ $A=1$, $\lambda =10.$ Also (a) and (b) are correspondence with $%
V(r,\phi )=V_{+}$ and $V_{c}$ respectively.

Figure (7): A plot of probability density $\left\vert \Psi _{N}\left( r,\phi
\right) \right\vert ^{2}$ for $N=7$ (i.e. $2\leq l\leq 9)$, $\theta _{\circ
}=0,$ $A=1$, $\lambda =100.$

Figure (8): A plot of $\left\vert I_{2l}\left( 2\lambda _{1}\cos \left( 
\frac{\phi }{2}\right) +i2\lambda _{2}\sin \left( \frac{\phi }{2}\right)
\right) \right\vert ^{2}$ in terms of $\phi ,$ for some different values of $%
\lambda _{1}$ and $\lambda _{2}.$

Figure (9): A plot of $\left\vert I_{2l}\left( 2\lambda _{1}\cos \left( 
\frac{\phi }{2}\right) +i2\lambda _{2}\sin \left( \frac{\phi }{2}\right)
\right) \right\vert ^{2}$ in terms of $\phi ,$ for some different values of $%
\lambda _{1}$ and $\lambda _{2}.$

Figure (10): A plot of $\left\vert I_{2l}\left( 2\lambda _{1}\cos \left( 
\frac{\phi }{2}\right) +i2\lambda _{2}\sin \left( \frac{\phi }{2}\right)
\right) \right\vert ^{2}$ in terms of $\phi ,$ for some different values of $%
\lambda _{1}$ and $\lambda _{2}.$

\bigskip 

\end{document}